\documentstyle[12pt]{article}
\def\I{\leavevmode\hbox{\small1\kern-3.8pt\normalsize1}}

\newcommand{\be}{\begin{equation}}
\newcommand{\ee}{\end{equation}}
\newcommand{\bq}{\begin{eqnarray}}
\newcommand{\eq}{\end{eqnarray}}
\newcommand{\bqc}{\begin{eqnarray*}}
\newcommand{\eqc}{\end{eqnarray*}}

\def\llsymbol#1{\@llsymbol{\@nameuse{c@#1}}}
\def\@llsymbol#1{\ifcase#1\or {}\or {'}\or {''}\or {'''}\or
   {''''}\or {'''''}\or  \else\@ctrerr\fi\relax}

\newcounter{contador}
\newcommand{\letra}{
   \stepcounter{equation}
   \setcounter{contador}{\value{equation}}
   \setcounter{equation}{0}
   \renewcommand{\theequation}{\thecontador.\alph{equation}}}
\newcommand{\antiletra}{
   \renewcommand{\theequation}{\arabic{equation}}
   \setcounter{equation}{\value{contador}}}

\begin{document}
\title{Some Quantum Aspects of Complex Vector Fields with Chern-Simons Term}
\author{O.M. Del Cima\thanks{Bitnet e-mail: oswaldo@brlncc.bitnet}
\thanks{Internet e-mai: oswaldo@cbpfsu1.cat.cbpf.br} \\
Brazilian Centre for Physical Researches (CBPF) \\
Department of Fields and Particles (DCP) \\
Rio de Janeiro - RJ - Brazil.
\and F.A.B. Rabelo de Carvalho \\
Catholic University of Petr\'opolis (UCP) - ICEN \\
Petr\'opolis - RJ - Brazil.}

\date{}

\maketitle

\begin{abstract}
Complex vector fields with Maxwell, Chern-Simons and Proca terms are minimally
coupled to an Abelian gauge field. The consistency of the spectrum is analysed
and 1-loop quantum corrections to the self-energy are explicitly computed and
discussed. The incorporation of 2-loop contributions and the behaviour of
tree-level scattering amplitudes in the limit of high center-of-mass energies
are also commented.
\end{abstract}

\newpage

\section{Introduction}

One of the central problems in the framework of gauge field theories is the
issue of gauge field mass. Gauge symmetry is not, in principle, conflicting
with the presence of a massive gauge boson. In 2 space-time dimensions, the
well-known Schwinger model puts in evidence the presence of a massive photon
without the breaking of gauge symmetry {\cite{1}}: a dynamical mass generation
takes place by virtue of fermion 1-loop corrections to the Maxwell field
polarization tensor.

Another evidence for the compatibility between gauge symmetry and massive
vector fields comes from the study of 3-dimensional gauge theories
{\cite{2,3}}. A topological mass term referred to as the Chern-Simons
Lagrangian, once added to the Maxwell kinetic term, shifts the photon mass to a
non-vanishing value without breaking gauge invariance at all {\cite{2,3}}. Even
if the Chern-Simons term, which is gauge invariant, is not written down at
tree-level, it may be generated by 1-loop corrections whenever massive fermions
are minimally coupled to an Abelian gauge field {\cite{4,5,6}}. Again, a
dynamical mass generation mechanism takes place. Also, in 3 space-time
dimensions, there occurs a dynamical fermionic mass generation if massless
fermions are minimally coupled to a Chern-Simons field {\cite{4,5,6,7}}.

In the more realistic case of 4 space-time dimensions, the best mechanism
known, up to now, to solve the problem of intermediate boson masses is the
spontaneous symmetry breaking mechanism {\cite{8,9}}. It is not known any
4-dimensional counterpart of the dynamical mechanism to generate gauge field
masses along the lines previously mentioned. However, in 4 dimensions, one
should quote the dynamical breaking of chiral symmetry which takes place
through a dynamical mass generation mechanism for fermions {\cite{10,11}}.

Since, over the past years, 3-dimensional field theories have been shown to
play a central r\^ole in connection with the behaviour of 4-dimensional
theories at finite temperature {\cite{12}} and in the description of a number
of problems in Condensed Matter Physics {\cite{13}}, it seems reasonable to
concentrate efforts in trying to understand some peculiar features of gauge
field dynamics in 3 dimensions. Also, the recent result on the Landau gauge
finiteness of Chern-Simons theories is a remarkable property that makes
3-dimensional gauge theories so attractive {\cite{finiteness}}.

The main purpose of this paper is to consider 3-dimensional models built up in
terms of complex vector fields with Chern-Simons terms and to which one
minimally couples a Maxwell field. At tree-level, we study the
Chern-Simons-Maxwell (CSM$^*$) and the Chern-Simons-Maxwell-Proca (CSMP$^*$)
cases, in order to analyse the conditions to be set on the free parameters of
the Lagrangians, so as to avoid the presence of tachyons and ghosts in the
spectrum. This is carried out in Section 2. In Section 3, we study the Abelian
CSM$^*$ model and show that, upon the incorporation of 1-loop corrections to
the CSM$^*$-field self-energy, a finite Proca mass term is generated. The
analysis of Section 2, in combination with the latter result, ensures that the
generated Proca-like term does not plug the theory with tachyons or ghosts.
Finally, in Section 4, we discuss the incorporation of 2-loops contributions to
the model, some results concerning the behaviour of the scattering amplitudes
in the limit of very high center-of-mass energies are discussed and we draw our
general conclusions. One Appendix follows where the explicit results for the
momentum-space 1-loop integrals are collected. The metric adopted throughout
this work is $\eta_{\mu \nu}=(+;-,-)$.

\section{The Complex Chern-Simons-Maxwell (CSM$^*$) and
Chern-Simons-Maxwell-Proca (CSMP$^*$) Fields}

The CSM$^*$ model is described by the Lagrangian

\begin{equation}
{\cal L}^0_{CSM}={1 \over 2} \epsilon^{\alpha \mu \nu} B^*_{\alpha} G_{\mu \nu}
- {1 \over 2M} G^*_{\mu \nu} G^{\mu \nu} \;\;, \label{2.1}
\end{equation}
where $G_{\mu \nu}$$\equiv$$\partial_{\mu}B_{\nu} $$-$$\partial_{\nu}B_{\mu}$
and
$G^*_{\mu \nu}$$\equiv$$\partial_{\mu}B^*_{\nu} $$-$$\partial_{\nu}B^*_{\mu}$
are the field-strengths, and $M$ is a real parameter with dimension of mass.

There are two kinds of $U(1)$ symmetries that may be observed in (\ref{2.1}). A
global $U_{\alpha}(1)$ given by
\begin{equation}
B'_{\mu}(x) = e^{i\alpha} B_{\mu}(x) \;\;, \label{2.2}
\end{equation}
where $\alpha$ is a real parameter, and a local $U_{\beta}(1)$ that reads
\begin{equation}
B'_{\mu}(x) = B_{\mu}(x) + \partial_{\mu}\beta(x) \;\;, \label{2.3}
\end{equation}
where $\beta(x)$ is an arbitrary $C^\infty$ complex function. The question
involving gauge symmetries with complex parameters has already been
contemplated in the context of spontaneously broken symmetries in
supersymmetric gauge models {\cite{14}}.

To minimally couple the CSM$^*$ fields, $B_{\mu}$ and $B^*_{\mu}$, to the
Maxwell field, $A_{\mu}$, we define the following $U_{\alpha}(1)$-covariant
derivatives :
\begin{equation}
D_{\mu}\equiv \partial_{\mu} + i \omega A_{\mu}\;\;\mbox{and}\;\;
D^*_{\mu}\equiv \partial_{\mu} - i \omega A_{\mu} \;, \label{2.4}
\end{equation}
where $\omega$ is a coupling constant with dimension of (mass)$^{1\over2}$.
Then,
the total Lagrangian becomes
\begin{eqnarray}
{\cal L}_{CSM}(B,B^*,\partial{B},\partial{B^*},A) &=& {\cal
L}^0_{CSM}(B,B^*,DB,D^*B^*) -
{1\over4}F_{\mu \nu}F^{\mu \nu}  \nonumber \\
&=& {1\over2}\epsilon^{\alpha \mu \nu} B^*_{\alpha} \tilde{G}_{\mu \nu} -
{1\over2M} \tilde{G}^*_{\mu \nu} \tilde{G}^{\mu \nu} -
{1\over4} F_{\mu \nu}F^{\mu \nu} \;, \label{2.5}
\end{eqnarray}
where $\tilde{G}_{\mu \nu}$$\equiv $$D_{\mu}B_{\nu}$$-$$D_{\nu}B_{\mu}$, and
$F_{\mu \nu}$ is the field-strength for $A_{\mu}$. By replacing the covariant
derivatives as given in eq.(\ref{2.4}), the total Lagrangian reads :
\begin{eqnarray}
{\cal L}_{CSM}&=&-{1\over4}F_{\mu \nu}F^{\mu \nu}+{1\over2}\epsilon^{\alpha \mu
\nu}
B^*_{\alpha}G_{\mu \nu}-{1\over2M}G^*_{\mu \nu}G^{\mu \nu}+i\omega
\epsilon^{\alpha \mu \nu}
B^*_{\alpha}A_{\mu}B_{\nu}+  \nonumber \\
&&-i{\omega \over M}(G^{*}_{\mu \nu}A^{\mu}B^{\nu}-G_{\mu \nu}A^{\mu}B^{* \nu})
-
{\omega^{2} \over M}(A_{\mu}B_{\nu}-A_{\nu}B_{\mu})A^{\mu}B^{* \nu} \;.
\label{2.6}
\end{eqnarray}
It can be noticed that the local $U_{\beta}(1)$-symmetry (\ref{2.3}) is
explicitly broken by the interaction terms in (\ref{2.6}).

In order to perform the analysis of the spectral consistency of this
model, it is necessary to obtain the propagator
for the fields $B$ and $B^*$. Since the local $U_{\beta}(1)$-symmetry  is
broken only at the interaction level, we need a gauge-fixing term to be able to
read off the propagators. So, for the sake of extracting them, we consider the
Lagrangian below :
\begin{equation}
\hat{\cal L}^0_{CSM} = {1 \over 2} \epsilon^{\alpha \mu \nu} B^{*}_{\alpha}
G_{\mu \nu} - {1 \over 2M} G_{\mu \nu}^{*} G^{\mu \nu} + {1 \over \hat
\alpha}(\partial_{\mu} B^{* \mu}) (\partial_{\nu} B^{\nu})\;, \label{2.7}
\end{equation}
where $\hat \alpha$ is the gauge-fixing parameter.

The field equations coming from (\ref{2.7}) are given by
\begin{equation}
{\cal O}^{\epsilon \alpha} B^{*}_{\alpha} = 0\;, \label{2.8}
\end{equation}
with
\begin{equation}
{\cal O}^{\epsilon \alpha} \equiv - \epsilon^{\epsilon k \alpha} \partial_k -
{\Box \over M} \left(\eta^{\epsilon \alpha} - {\partial^{\epsilon}
\partial^{\alpha} \over \Box}\right) +
{\Box \over \hat \alpha} \left({\partial^{\epsilon} \partial^{\alpha} \over
\Box}\right)\;,
			\label{2.9}
\end{equation}
where
\begin{equation}
{\Theta}^{\mu \nu} \equiv \eta^{\mu \nu} -
{\partial^{\mu}\partial^{\nu}\over\Box}\;,\;\;S^{\mu \nu} \equiv
\epsilon^{\mu \alpha \nu}
\partial_{\alpha}\;\;\mbox{and}\;\;\Omega^{\mu \nu} \equiv
{\partial^{\mu} \partial^{\nu} \over \Box}
\end{equation}
are spin operators that fulfil the algebra displayed in Table 1.
\begin{table}
\label{tab}
\begin{center}
\begin{tabular}{|c|r|r|r|} \hline
			& $\;\;\Omega\;\;$ & $\;\;\Theta\;\;$ & $\;\;S\;\;$
\\ \hline
$\;\;\Omega\;\;$ & $\;\;\Omega\;\;$ & $\;\;\bf{0}\;\;$ & $\;\;\bf{0}\;\;$   \\
\hline
$\;\;\Theta\;\;$ & $\;\;\bf{0}\;\;$ & $\;\;\Theta\;\;$ & $\;\;S\;\;$        \\
\hline
$\;\;S\;\;$      & $\;\;\bf{0}\;\;$ & $\;\;S\;\;$      & $-\;\Box \;\Theta$ \\
\hline
\end{tabular}
\end{center}
\caption{Operator algebra fulfilled by $\Omega$,$\Theta$ and $S$.}
\end{table}

Inverting the operator ${\cal O}$ with the help of the Table 1, we obtain the
following momentum-space propagators in the longitudinal and transverse
subspaces, respectively :
\letra
\begin{eqnarray}
\Delta^{\mu \nu}_{L} (k) &=& i\; {\hat \alpha \over k^2} \left({k^{\mu}
k^{\nu} \over k^2} \right) \label{2.11a} \\
\noalign{\hbox{and}}\nonumber \\
\Delta^{\mu \nu}_{T} (k) &=& -i\; {M^2 \over k^2(k^2-M^2)}
\left[\;i\;\epsilon^{\mu k \nu}
k_{k} + {k^2 \over M} \left(\eta^{\mu \nu} - {k^{\mu} k^{\nu} \over
k^2} \right) \right]\;. \label{2.11b}
\end{eqnarray}
\antiletra

By saturating the propagators with external conserved currents, $J^{\mu}$ and
$J^{\mu *}$, the following result on the spectrum can be stated :
\letra
\be
L-sector \longrightarrow {\hbox{pole at}}\; k^2=0
\;{\hbox{non-dynamical}}  \label{2.12a}
\ee
\be
T-sector \longrightarrow \cases{ {\hbox{pole at}}\; k^2=0
\;{\hbox{non-dynamical}}
\cr {\hbox{pole at}}\; k^2=M^2 \cases{ {\hbox{dynamical}} \cr
{\hbox{no tachyons, no ghosts}} \;\; if \;\; M \; > \; 0} \cr}\;. \label{2.12b}
\ee
\antiletra

Thus, we may conclude that, once the mass parameter, $M$, is taken to be
positive, the CSM$^*$ model describes a free physical dynamical excitation of
mass $k^2=M^2$.

The CSMP$^*$ model is described by a Lagrangian obtained from (\ref{2.1}) by
the addition of a Proca
term, $\hat{\mu} B^*_\mu B^\mu$. Then,
\be
{\cal L}^o_{CSMP} = {1 \over 2} \epsilon^{\alpha \mu \nu} B^*_\alpha
G_{\mu \nu} - {1 \over 2M} G^*_{\mu \nu} G^{\mu \nu} + \hat{\mu} B^*_\mu
B^\mu \;, \label{3.1}
\ee
where $\hat{\mu}$ is a real parameter with mass dimension.


It may be observed that the Lagrangian of eq.(\ref{3.1})
exhibits only one global symmetry, $U_{\alpha}(1)$ :
\begin{equation}
B'_{\mu}(x) = e^{i\alpha} B_{\mu}(x) \;\;, \label{3.2}
\end{equation}
where $\alpha$ is a real parameter. The local symmetry $U_{\beta}(1)$
(\ref{2.3}) is explicitly broken by the Proca term.

Carrying out the minimal coupling of the CSMP$^*$ fields,
$B_{\mu}$ and $B^*_{\mu}$, to the Maxwell field $A_{\mu}$, one gets the
Lagrangian
\begin{eqnarray}
{\cal L}_{CSMP}&=&- {1 \over 4} F_{\mu \nu} F^{\mu \nu} + {1 \over 2}
\epsilon^{\alpha \mu \nu} B^*_\alpha G_{\mu \nu} - {1 \over 2M} G^*_{\mu
\nu} G^{\mu \nu} + \hat{\mu} B^*_\mu B^\mu + \nonumber \\
&&+i \omega \epsilon^{\alpha
\mu \nu} B^*_\alpha A_\mu B_\nu - i {\omega \over M} ( G^*_{\mu \nu}
A^\mu B^\nu - G_{\mu \nu} A^\mu B^{*\nu}) + \nonumber \\
&&-{\omega^2 \over M} (A_\mu B_\nu
- A_\nu B_\mu ) A^\mu B^{* \nu} \;.         \label{3.3}
\end{eqnarray}

To pursue our investigation on the consistency of the spectrum, we shall now
quote the expressions derived for the propagators of the CSMP$^*$ fields and
then analyse their poles and associated residues.

The momentum-space expressions for the propagators are :
\letra
\begin{eqnarray}
\overline \Delta^{\mu \nu}_{L} (k)&=& i\; {1 \over \hat{\mu}}
\left({k^\mu k^\nu \over k^2} \right)  \label{3.7a}\\
\noalign{\hbox{and}}\nonumber \\
\overline \Delta^{\mu \nu}_{T} (k)&=& -i\; {M \over [(k^2 - \hat{\mu}
M)^2 - M^2 k^2]} \left[ i M \epsilon^{\mu \kappa \nu} k_\kappa + (k^2 -
\hat{\mu} M) \left( \eta^{\mu \nu} - {k^\mu k^\nu \over k^2} \right)
\right]\nonumber \\
 &=& -i\; {M \over (k^2 - m^2_+)(k^2 - m^2_-)} \biggl[i M
\epsilon^{\mu \kappa \nu} k_\kappa + (k^2 - \hat{\mu} M) \left(
\eta^{\mu \nu} - {k^\mu k^\nu \over k^2} \right) \biggr]
\;,\label{3.7b}\nonumber \\
\end{eqnarray}
\antiletra
where
\letra
\bq
m^2_+&\equiv&{M \over 2}\; [ M + 2 \hat{\mu} + \sqrt{M(M+4 \hat{\mu})}\ ]
\label{3.8a}\\
\noalign{\hbox{and}}\nonumber \\
m^2_-&\equiv&{M \over 2}\; [M + 2 \hat{\mu} -
\sqrt{M(M+ 4 \hat{\mu})} \ ]\;, \label{3.8b}
\eq
\antiletra
with $M(M+4\hat{\mu})\geq 0$, in order to avoid unphysical complex roots.

To avoid the apperance of a double pole, $m^2_+=m^2_-$ (which would certainly
lead to a ghost), we must actually have $M(M+4\hat{\mu}) > 0$.

Again, we saturate the propagators with external conserved currents, and the
following results on the spectrum hold :
\be
T-sector \longrightarrow \cases{{ {\hbox{pole at}}\; k^2=m^2_+
{\cases{{\hbox{dynamical}} \cr  {\hbox{no tachyons, no ghosts}} \;\;if\;\;  M
\; {\rm and} \; \hat{\mu} >0 }}\\} \cr
{\hbox{pole at}}\; k^2=m^2_-
\cases{ {\hbox{dynamical}} \cr
{\hbox{no tachyons, no ghosts}} \;\;if\;\;
  M \; {\rm and} \; \hat{\mu} >0  \cr}}\;.\label{3.9b}
\ee

The analysis of the residues shows that the $T$-sector is free from tachyons
and ghosts whenever $\hat{\mu}>0$ and $M>0$.

Also, the conditions $\hat{\mu}>0$ and $M>0$ automatically avoid a double pole.
Then, the CSMP$^*$ model is perfectly physical, as long as the spectrum is
concerned, if these two conditions are set.

Nevertheless, to control the issue of unitarity at tree-level, it is necessary
to study the behaviour of scattering cross sections in the limit of very high
center-of-mass energies. This has been discussed in detail in the paper of
ref.{\cite{15}}, where we have illustrated that the scattering process between
a CSM$^*$ vectorial boson and the photon exhibits a cross-section whose
asymptotic behaviour respects the Froissart bound in 3 dimensions.

A peculiar feature concerns the presence of two different simple poles in the
transverse sector of the propagator for the CSMP$^*$-field. This is also a
characteristic of a real CSMP-field. The poles are to be interpreted as two
distinct excitations whose spins have to be fixed in terms of the masses, after
a detailed analysis of the Lorentz group generators as functionals of the
fields is carried out, in the same way it is done for a topologically massive
theory {\cite{3}}. However, each of the masses has a definite value for the
spin, $\pm 1$, since there is no room for different polarization states in
$D=3$. Hence, the 2 degrees of freedom of the real CSMP-field correspond to the
2 possible values for the mass, $m_{\pm}^2$ . In the complex case, the 4
degrees of freedom are associated to the 2 different states of charge that each
massive pole may present.

\section{Dynamical Mass Generation in the CSM$^*$ Model}

By reconsidering the Lagrangian (\ref{2.6}), the following interaction vertices
(see Fig.1) come out :
\letra
\begin{eqnarray}
{\cal L}^{(1)\;int}_{CSM} &=& i\; \omega \epsilon^{\alpha \mu \nu}\;
B^*_{\alpha}
A_{\mu} B_{\nu} \;\;\longrightarrow \;\;V_3 \; ,\\
{\cal L}^{(2)\;int}_{CSM} &=& -i\; {\omega \over M}\; (G^*_{\mu \nu}
A^{\mu} B^{\nu} - G_{\mu \nu} A^{\mu} B^{* \nu}) \;\;\longrightarrow
\;\;\overline V_3 \\
\noalign{\hbox{and}} \nonumber\\
{\cal L}^{(3)\;int}_{CSM} &=& -\;{\omega^{2} \over M} \;(A_{\mu} B_{\nu} -
A_{\nu}
B_{\mu}) A^{\mu} B^{* \nu} \;\;\longrightarrow \;\;V_4 \; .
\end{eqnarray}
\antiletra

Before the calculation of the Feynman graphs relevant for our analysis on the
mass generation, we present the expression we get for the superficial degree of
divergence of the primitively divergent graphs of the model.

Analysing the CSM$^*$ propagator in the high energy limit, and taking into
account the interaction vertices above, we find the following expression for
the superficial degree of divergence, $\delta_{CSM}$ :
\be
\delta_{CSM} = 3 - {3 \over 2}v_3 - {1 \over 2} \overline v_3 - v_4
-{1 \over 2} E_A - {1 \over 2} E_B \;, \label{4.2}
\ee
where $v_3$, $\overline v_3$ and $v_4$ are the numbers of vertices $V_3$,
$\overline V_3$ and $V_4$ respectively, $E_A$ are the external lines of
$A_{\mu}$ and $E_B$ are the external lines of $B_{\mu}$ and $B_{\mu}^{*}$.
Therefore, the CSM$^*$ is a super-renormalizable model: ultraviolet divergences
appear only up to 2-loops. Now, since in 3 space-time dimensions no 1-loop
divergences show up, all renormalizations have to be performed at 2-loops.

The vertex Feynman rules of the model read as below :
\letra
\bq
(V_3)_{\alpha \mu \nu} &=& \omega \epsilon_{\alpha \mu \nu} \;\;,\label{4.3a}\\
(\overline V_3)_{\alpha \mu \nu} &=& i {\omega \over M} (\eta_{\nu \alpha}
k_{\mu} -
\eta_{\mu \alpha} k_{\nu} + \eta_{\nu \alpha} m_{\mu} - \eta_{\mu \nu}
m_{\alpha}) \label{4.3b}\\
\noalign{\hbox{and}} \nonumber \\
(V_4)_{\alpha \nu \beta \mu} &=& i {2 \omega^2 \over M} (\eta_{\alpha \beta}
\eta_{\mu \nu} - \eta_{\alpha \nu} \eta_{\beta \mu}) \;.  \label{4.3c}
\eq
\antiletra

In Fig.2, we list the 1-loop diagrams that contribute to the CSM$^*$-field
self-energy. The explicit results for these diagrams may be found in
ref.{\cite{tese}}, where computations have been carried out in Landau gauge,
$\hat \alpha$$=$$0$.

Bearing in mind that we are concerned with the possibility of inducing a 1-loop
(finite) mass contribution of the Proca type, we can select only those terms
that do not exhibit any dependence on the external momenta and are moreover
symmetric on the free indices of the external lines. Therefore, the only terms
that potentially contribute a finite Proca mass term have been found to be
given by the following parametric integrals :
\letra
\bq
(I_1)_{\alpha \beta}&=& {\omega^2 \over M} \int {d^3k \over (2 \pi)^3}
{k_\alpha k_\beta \over (k-p)^2(k^2-M^2)}\nonumber \\ &=& i {\omega^2 \over 8
\pi M} \bigg \{ p_\alpha p_\beta
\int^1_0 dx {x^2 \over [p^2x^2-(p^2+M^2)x+M^2]^{1 \over 2}} + \nonumber \\
&&+\;\eta_{\alpha \beta} \int^1_0 dx [p^2 x^2 - (p^2 + M^2)x + M^2]^{1 \over 2}
\bigg \} \; ,\label{4.5a}\\
(I_2)_{\alpha \beta}&=& {\omega^2 \over M} \eta_{\alpha \beta} \int
{d^3k \over (2 \pi)^3} {k^2 \over (k-p)^2(k^2-M^2)}\nonumber \\
&=& i {\omega^2 \over 8 \pi M} \eta_{\alpha \beta} \bigg
\{ p^2 \int^1_0 dx {x^2 \over [p^2x^2-(p^2+M^2)x+M^2]^{1 \over 2}} + \nonumber
\\
&&+\;3 \int^1_0 dx [p^2 x^2 - (p^2 + M^2)x +M^2]^{1 \over 2} \bigg \} \;
,\label{4.5b}\\
(I_3)_{\alpha \beta}&=& M \omega^2 \eta_{\alpha \beta} \int {d^3k \over
(2 \pi)^3} {1 \over (k-p)^2(k^2-M^2)}\nonumber \\
&=& i {\omega^2 M \over 8 \pi} \eta_{\alpha \beta}
\int^1_0 dx {1 \over [p^2 x^2- (p^2 + M^2 )x + M^2]^{1 \over 2}} \label{4.5c}\\
\noalign{\hbox{and}} \nonumber\\
(I_4)_{\alpha \beta} &=& M \omega^2 \int {d^3k \over (2 \pi)^3} {k_\alpha
k_\beta \over (k-p)^2(k^2-M^2)k^2}\nonumber \\
&=& i {\omega^2 \over 8 \pi M} \bigg \{ p_\alpha p_\beta
\int^1_0 dx {x^2 \over [p^2 x^2 - (p^2 + M^2)x + M^2]^{1 \over 2}} +
\nonumber\\
&&+\;\eta_{\alpha \beta} \int^1_0 dx [p^2x^2-(p^2+M^2)x + M^2]^{1 \over 2} +
\nonumber\\
&&-\; p_\alpha p_\beta \int^1_0 dx {x^2 \over (p^2 x^2 - p^2 x)^{1 \over 2}} -
\eta_{\alpha \beta} \int^1_0 dx (p^2 x^2- p^2x)^{1 \over 2} \bigg
\}\;.\label{4.5d}
\eq
\antiletra

Their explicit results are presented in the Appendix. By observing these
results (see \ref{B11}, \ref{B12}, \ref{B13} and \ref{B14}), we conclude that a
1-loop term given by $i {\omega^2 \over
32 \pi} \eta_{\alpha \beta}$, coming from $I_1$ and $I_4$, will lead to the
generation of the Proca term.

The whole 1-loop CSM$^*$ self-energy diagram, $\Omega^{(1)}$, is the sum of the
diagrams  $\Sigma$, $\Lambda$, $\Xi^{R}$, $\Xi^{L}$ and $\Gamma$ of Fig.2 :
\be
\Omega^{(1)} = \Sigma + \Lambda + \Xi^{R} + \Xi^{L} +\Gamma \;.\label{4.6}
\ee

By summing up all these pieces, we finally get that the 1-loop induced Proca
term comes from the contribution
\be
\Omega^{(1) \alpha \beta}_{\hat{\mu}} = i \;{\omega^2 \over 8 \pi}
\;\eta^{\alpha
\beta} = i \;\hat{\mu}\; \eta^{\alpha \beta} \label{4.7}\;,
\ee
from which we can readily read the Proca mass :
\be
\hat{\mu} = {\omega^2 \over 8 \pi} > 0 \label{4.8}\;.
\ee

It is interesting to emphasize that the term $\Omega^{(1)}_{\hat{\mu}}$,
generated by the 1-loop quantum corrections to the CSM$^*$ self-energy, is a
finite one, therefore it will not be necessary to add any counter-term to the
Lagrangian (\ref{2.6}). Such a finite term amounts to the contribution
\be
{\cal L}^{(1)}_{\hat{\mu}}=\hat{\mu} B^*_\mu B^\mu  \label{4.9}
\ee
to the classical Lagrangian. Since the parameter ${\hat{\mu}}$ in ${\cal
L}^{(1)}_{\hat{\mu}}$ automatically satisfies the condition ${\hat{\mu}}>0$,
the espectral consistency discussed in the previous section is not jeopardized.

\begin{figure}
\vspace{7cm}
\center{Figure 1: Interaction 3- and 4-vertices, $V_3$, $\overline V_3$ and
$V_4$.}
\label{fig1}
\end{figure}

\begin{figure}
\vspace{8cm}
\center{Figure 2: 1-loop CSM$^*$-field self-energy diagrams.}
\label{fig2}
\end{figure}

\section{Discussions and General Conclusions}

Our basic proposal in this paper has been to understand a number of features
concerning the dynamics of complex vector fields in $D$$=$$1$$+$$2$.

The first step of our study consisted in establishing conditions under which a
general CSMP complex vector field describes physically acceptable excitations.
It was obtained that such a complex vector field describes, in principle, two
distinct massive excitations, each of them appearing of course in two states
with opposite charges.

Having understood how to control the physical character of the quanta of the
model, we proposed to study the dynamics of a CSM$^*$-field minimally coupled
to an Abelian vector field (Maxwell field). The explicit calculation of 1-loop
corrections revealed the generation of a (finite) Proca term that was not
present at tree-level, respecting the spectral conditions set on the study of
the propagation of the CSMP$^*$-field. We then concluded that the 1-loop Proca
mass generation does not introduce neither tachyons nor ghosts in the spectrum.

The study concerning the behaviour of the ``Compton'' scattering cross sections
in the limit of very high (much higher than the masses of the quanta)
center-of-mass energies revealed that the CSM$^*$-model respects the Froissart
bound for $D$$=$$1$$+$$2$, while the CSMP$^*$-model violates this bound.
Moreover, contrary to what happens in the case of 4-dimensional massive charged
vector fields coupled to the Maxwell-field, Froissart bound cannot be restored
at the expenses of a gauge-invariant non-minimal coupling {\cite{15}}.

Also, another delicate point should be discussed. The CSM$^*$-model pre\-sents
divergences at the 2-loop level. Therefore, it is crucial to check whether or
not a ultraviolet divergent term of the form $|(\partial_{\mu}B^{\mu})|^2$
appears as a 2-loop contribution to the CSM$^*$-field self-energy. In view of
this result, one may have to add, for the sake of renormalization, the term
$|(\partial_{\mu}B^{\mu})|^2$ already at the classical level, and ghosts will
unavoidably show up that spoil the spectrum {\cite{priv}}. To clarify this
matter, one has to investigate the 2-loop self-energy graphs for the
CSM$^*$-field displayed in Fig.3.

Nevertheless, based on the power-counting ({\ref{4.2}}) derived for the model
we are considering, we find out that the graphs drawn in Fig.3 that involve
exclusively $\overline V_3$ and $V_4$ vertices are all logarithmically
divergent ($\delta_{CSM}$$=$$0$). On the other hand, since a
$p^{\mu}p^{\nu}$-dependence has to be factored out from these graphs so as to
build up a 2-loop correction of the form $|(\partial_{\mu}B^{\mu})|^2$, this
sort of contribution will consequently come out as a ultraviolet finite
correction to the effective action. This in turn means that no such a term,
which would for sure bring about longitudinal-mode ghosts, needs to be adjoined
to the tree-level action. Therefore, since we know (based on power-counting)
that from 3 loops on the model is totally finite, we can state that, though
gauge-invariant, a term like $|(D_{\mu}B^{\mu})|^2$ is not radiatively induced
into the effective action. Therefore, the spectral conditions established in
Section 2 are not spoiled whenever loop corrections are taken into account.

\begin{figure}
\vspace{12.25cm}
\center{Figure 3: 2-loop CSM$^*$-field self-energy logarithmically divergent
diagrams.}
\label{fig3}
\end{figure}

\section*{Appendix: 1-loop integrals for the CSM$^*$-field self-energy}

To solve the integrals $I_1$, $I_2$, $I_3$ and $I_4$ in Section 3, use has been
made of the following well-known results {\cite{Collins}} :
\bq
J_0 &=& \int {d^D k \over (2 \pi)^D} {1 \over (k^2 + 2p.k - c)^\alpha} =
i(-1)^\alpha {\pi^{D \over 2} \over (2\pi)^D} (c+p^2)^{{D \over 2} -
\alpha}\times \nonumber \\
&&\times {\Gamma(\alpha - {D \over 2}) \over \Gamma(\alpha)} \;  ,\label{b1}\\
\nonumber \\
J_1^\mu &=& \int {d^D k \over (2 \pi)^D} {k^{\mu} \over (k^2 + 2p.k -
c)^\alpha} =
i(-1)^{\alpha + 1} {\pi^{D \over 2} \over (2\pi)^D} (c+p^2)^{{D \over 2} -
\alpha} \times \nonumber \\
&&\times\; p^{\mu} {\Gamma(\alpha - {D \over 2}) \over \Gamma(\alpha)}
\label{b2}\\
\noalign{\hbox{and}} \nonumber \\
J_2^{\mu \nu} &=& \int {d^D k \over (2 \pi)^D} {k^{\mu} k^{\nu} \over (k^2 +
2p.k - c)^\alpha} = i(-1)^\alpha {\pi^{D \over 2} \over (2\pi)^D}
(c+p^2)^{{D \over 2} - \alpha} \times \nonumber \\
&&\times \left[ {\Gamma(\alpha - {D \over 2}) p^{\mu}
p^{\nu} - {1 \over 2} \Gamma \left(\alpha - 1 -  {D \over 2} \right) \eta^{\mu
\nu} (c+p^2) \over \Gamma(\alpha)} \right] \;. \label{b3}
\eq

We quote in the sequel the results for the Feynman parametric integrals
performed after the integration over the loop momenta :
\bq
(I_1)_{\alpha \beta}&=& \cases{ i{\omega^2 \over 8 \pi M} \biggl\{ p_\alpha
p_\beta \left[ - {3(p^2+M^2) \over 4(p^2)^2} M + {3 (p^2)^2 + 2M^2 p^2 + 3M^4
\over 8(p^2)^2} W \right]+  \cr
+\; \eta_{\alpha \beta} \left[ {M^3 \over 4p^2} - {(p^2 - M^2)^2 \over 8p^2} W
\right] \biggr\} +
i {\omega^2 \over 32 \pi} \eta_{\alpha \beta} \;\;,\;\; p^2>0  \cr  \cr
\label{B11}
i {\omega^2 \over 8 \pi M} \biggl\{ p_\alpha p_\beta \left[ - {3(p^2+M^2)
\over 4(p^2)^2} M + {3 (p^2)^2 + 2M^2 p^2 + 3M^4 \over 8(p^2)^2} V \right]+ \cr
+\; \eta_{\alpha \beta} \left[ {M^3 \over 4p^2} + {(p^2 - M^2)^2 \over 8p^2} V
\right] \biggr\} +
i {\omega^2 \over 32 \pi} \eta_{\alpha \beta} \;\;,\;\; p^2<0 \cr} \\
\nonumber \\
(I_2)_{\alpha \beta}&=& \cases{ i{\omega^2 M \over 8 \pi} \eta_{\alpha \beta} W
\;\;,\;\; p^2>0  \cr \label{B12}
i{\omega^2 \over 8 \pi M} \eta_{\alpha \beta} \left[ {3(p^2)^2 - 2M^2p^2+3M^4
\over 4p^2} V \right] \;\;,\;\; p^2<0 \cr} \\
\nonumber \\
(I_3)_{\alpha \beta}&=& \cases{ i{\omega^2 M \over 8 \pi} \eta_{\alpha \beta} W
\;\;,\;\; p^2>0  \cr \label{B13} i{\omega^2 M \over 8 \pi} \eta_{\alpha \beta}
V \;\;,\;\; p^2<0 \cr } \\
\nonumber \\
(I_4)_{\alpha \beta}&=& \cases{ i{\omega^2 \over 8 \pi M} \biggl\{ p_\alpha
p_\beta \left[ - {3(p^2 + M^2) \over 4(p^2)^2} M + i {3 \pi \over 8}
{1 \over \sqrt{p^2}} + {3(p^2)^2 + 2M^2 p^2+3M^4 \over 8(p^2)^2} W \right] +
\cr
+\; \eta_{\alpha \beta} \left[ {M^3 \over 4p^2} - i {\pi \over 8}
\sqrt{p^2} - {(p^2 - M^2)^2 \over 8p^2} W \right] \biggr\} +
i {\omega^2 \over 32 \pi} \eta_{\alpha \beta} \;\;,\;\; p^2>0  \cr \cr
\label{B14}
i{\omega^2 \over 8 \pi M} \biggl\{ p_\alpha p_\beta \left[ - {3(p^2 + M^2)
\over 4(p^2)^2} M - {3 \pi \over 8} {1 \over \sqrt{-p^2}} + {3(p^2)^2 +
2M^2 p^2+3M^4 \over 8(p^2)^2} V \right] + \cr
+\; \eta_{\alpha \beta} \left[{M^3 \over 4p^2} - {\pi \over 8} \sqrt{-p^2} +
{(p^2 - M^2)^2 \over 8p^2} V \right] \biggr\} + i {\omega^2 \over 32 \pi}
\eta_{\alpha \beta} \;\;,\;\; p^2<0 \cr}
\eq
where $W$ and $V$ are defined as
\be
W \equiv {1 \over \sqrt{p^2}} \left [ {\ln}( \vert p^2 - M^2 \vert ) - 2
{\ln} (\vert \sqrt{p^2} - M \vert) - i \pi \theta (p^2- M^2) \right ] \; \;
\mbox{if} \;\; p^2>0  \\
\ee
and
\be
V \equiv {1 \over \sqrt{-p^2}} \left [ {\pi \over 2} - {\arctan}
\left({p^2 + M^2 \over 2M \sqrt{-p^2}} \right) \right] \; \; \mbox{if} \;\;
p^2<0 \;.
\ee

\vspace{2cm}

\section*{Acknowledgements}

The authors express their gratitude to Dr. J.A. Helay\"el-Neto, Dr. O. Piguet,
Prof. M. Chaichian and Dr. S.P. Sorella for patient and helpful discussions.
Thanks are also due to Prof. H. Freitas de Carvalho and our colleagues at
CBPF-DCP. CNPq-Brazil and FAPERJ-Brazil are acknowledged for invaluable
financial help.

\end{document}